\pdfoutput=1 
\documentclass[biblatex]{lni}

\bibliography{bib/bibliography}

\usepackage{subfig}

\begin{document}
\title[Anomaly Detection using Graph Databases and Machine Learning]{Anomaly Detection in Log Data using Graph Databases and Machine Learning to Defend Advanced Persistent Threats}
\author[Timo Schindler]
{Timo Schindler\footnote{OTH Regensburg, Laboratory for Information Security, Regensburg, Germany \email{timo.schindler@othr.de}}}
\startpage{1} 
\editor{Gesellschaft für Informatik e.V.} 
\booktitle{Informatik 2017} 
\year{2017}
\maketitle

\begin{abstract}
Advanced Persistent Threats (APTs) are a main impendence in cyber security of computer networks. 
In 2015, a successful breach remains undetected 146 days on average, reported by \cite{fireeye:m-trends2016}.
With our work we demonstrate a feasible and fast way to analyse real world log data to detect breaches or breach attempts.
By adapting well-known kill chain mechanisms and a combine of a time series database and an abstracted graph approach, it is possible to create flexible attack profiles.
Using this approach, it can be demonstrated that the graph analysis successfully detects simulated attacks by analysing the log data of a simulated computer network.
Considering another source for log data, the framework is capable to deliver sufficient performance for analysing real-world data in short time.
By using the computing power of the graph database it is possible to identify the attacker and furthermore it is feasible to detect other affected system components.
We believe to significantly reduce the detection time of breaches with this approach and react fast to new attack vectors.
\end{abstract}

%
%
\section{Introduction}
Over the last 10 years, Vukalovic et al. show that a significant increase of Advanced Persistent Threats (APTs) on companies in all industries can be observed \cite{Vukalovic2015}.
The Verizon Data Breach Investigations Report \cite{verizon2016} claims that nearly 100\% of all compromises are happening in less than a day, where only about 10\% of those compromises are discovered in a day or less.
The Fireeye report \cite{fireeye:m-trends2016} also shows that security breaches of computer systems in 2015 remained undiscovered for 146 days in average.
In addition to that the recent security breach at thyssenkrupp reported by \cite{wiwo2016} and \cite{FinancialTimes}, with a time period of 45 days between breach and detection, shows that attackers have enough time to interact undetected in an intruded system.

The report of \cite{fireeye:m-trends2016} furthermore illustrates that only 47\% of the reported compromises are detected by an internal discovery.
Regarding APTs the attacker is aware of the defense mechanisms and will camouflage his actions.
Still every attack creates anomalies or changes the behaviour of user accounts which is represented in the raw log data.
To achieve a APT detection, all log data is analysed continuously, to create a profile, which represents regular behaviour, where no attack takes place.
Both, the profile creation and the consolidation of the raw log data is achieved with graph databases by using an adjusted kill chain model.
With this kill chain model it is possible to identify different attack vectors and increase the precision of the detection rate.
By introducing an abstracted event sequence layer, flexibility in detecting divergent attacks is added to the graph database model and is an efficient way to interpret multiple linked log events.
To create and test this attack vectors, a simulator is used to create malicious and clean (without attacks) raw log data.

%
%
\section{State of the Art}
Intrusion Detection or Inrustion Prevention Systems (IDS and IPS) are widespread approaches to detect security breaches or attacks (\cite{dell:ids,checkpoint:ids}).
Considering \cite{lin2015} IDS are developed for signature and/or anomaly detection.
For signature detection, packets or audit logs are scanned to look for sequences of commands or events which are previously determined as indicative of an attack.
Recently researchers are also working on innovative approaches including data mining, statistical analysis and artificial intelligence techniques (see \cite{lin2015}; \cite{1556540}; \cite{1408059}; \cite{INT:INT20203}).
Alongside IPS/IDS Security Information and Event Management (SIEM) is a widely used approach to gain security in companies and are also object of current research \cite{Yen:2013:BLL:2523649.2523670}.
\cite{lin2015} listed recent related work focusing on the task of anomaly detection based on various data mining and machine learning techniques.
Most studies are using the KDD-Cup 99 \cite{Rosset2000} or DARPA 1999 \cite{Lippmann2000} datasets.
In the field of research and also in existing solutions, data protection is no or just a minor issue.

%
%
\section{Objective}
The objective of this work is to create a feasible and efficient solution to analyse various sets of log data and detect APT attacks in a reliable way.
The graph database approach demonstrates a way to analyse multiple interrelated log events in a structured an straightforward process.
To achieve this, an abstraction layer is introduced, which is capable to react on divergent and looping log events.
By using a adapted kill chain model, different attack vectors can be identified.
This leads to a high level of confidence for the  case that several kill chain elements are met subsequently.
The created framework uses a state of the art SIEM system and widely used database solutions for fast data processing.
To comply to the German Federal Data Protection Act (\S3a BDSG\footnote{Bundesdatenschutzgesetz, Federal Data Protection Act}) all user related data has to be pseudonymised and the framework is able to to work with pseudonymised data without limitations on precision.
We proof that this work leads to a flexible way to detect anomalies from raw log data and is capable to be adjusted to several other Advanced Persistent Threats.

%
%
\section{Graph based Kill Chain Analysis}
\begin{figure}
\subfloat[Adjusted Kill Chain with an Example APT\label{fig:Adjusted-Kill-Chain}]{
\includegraphics[width=0.5\textwidth]{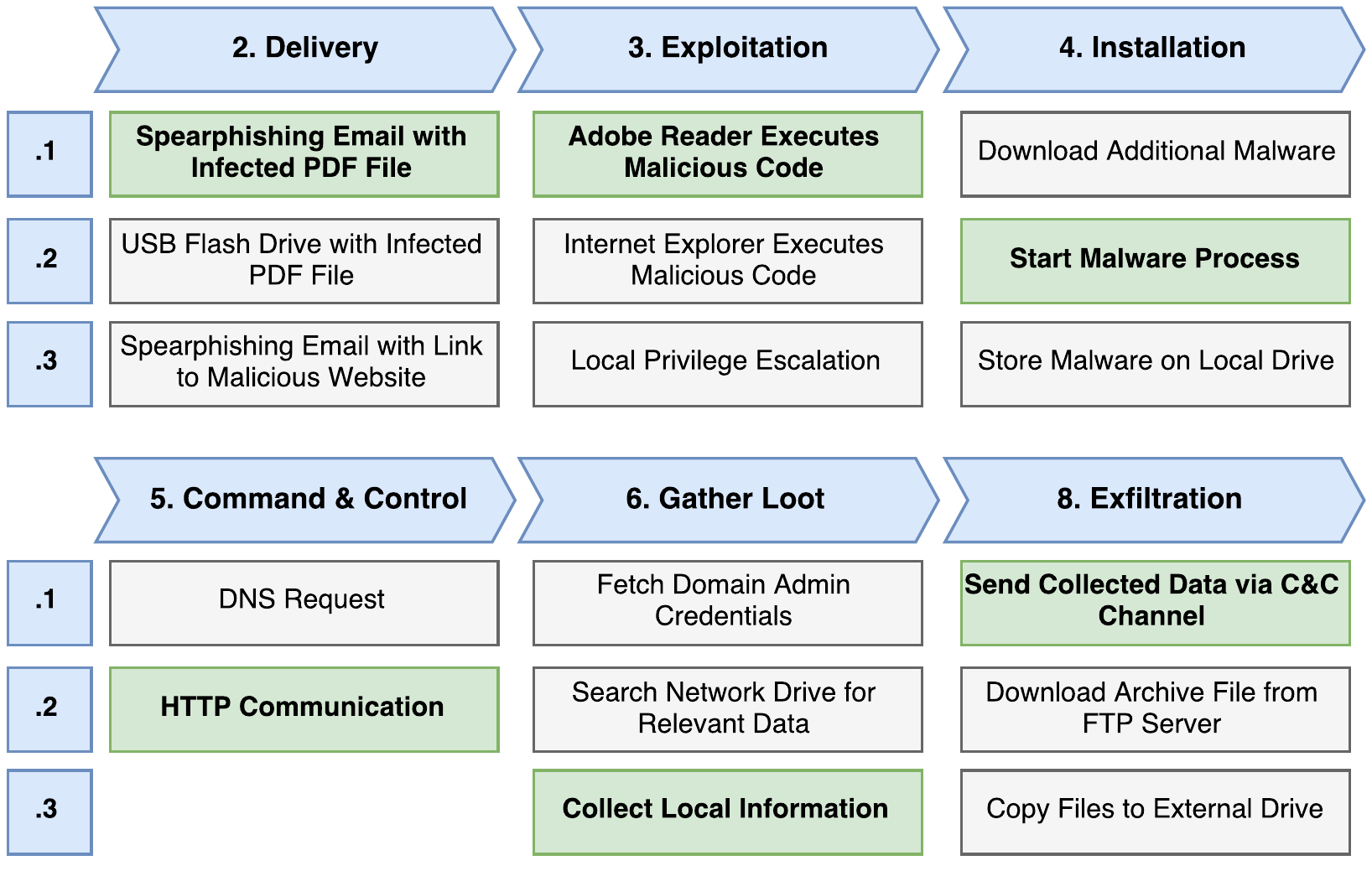}
\par
}\subfloat[Kill Chain Event Mapping with Event Sequences\label{fig:Kill-Chain-Event-Mapping}]{
\includegraphics[width=0.5\textwidth]{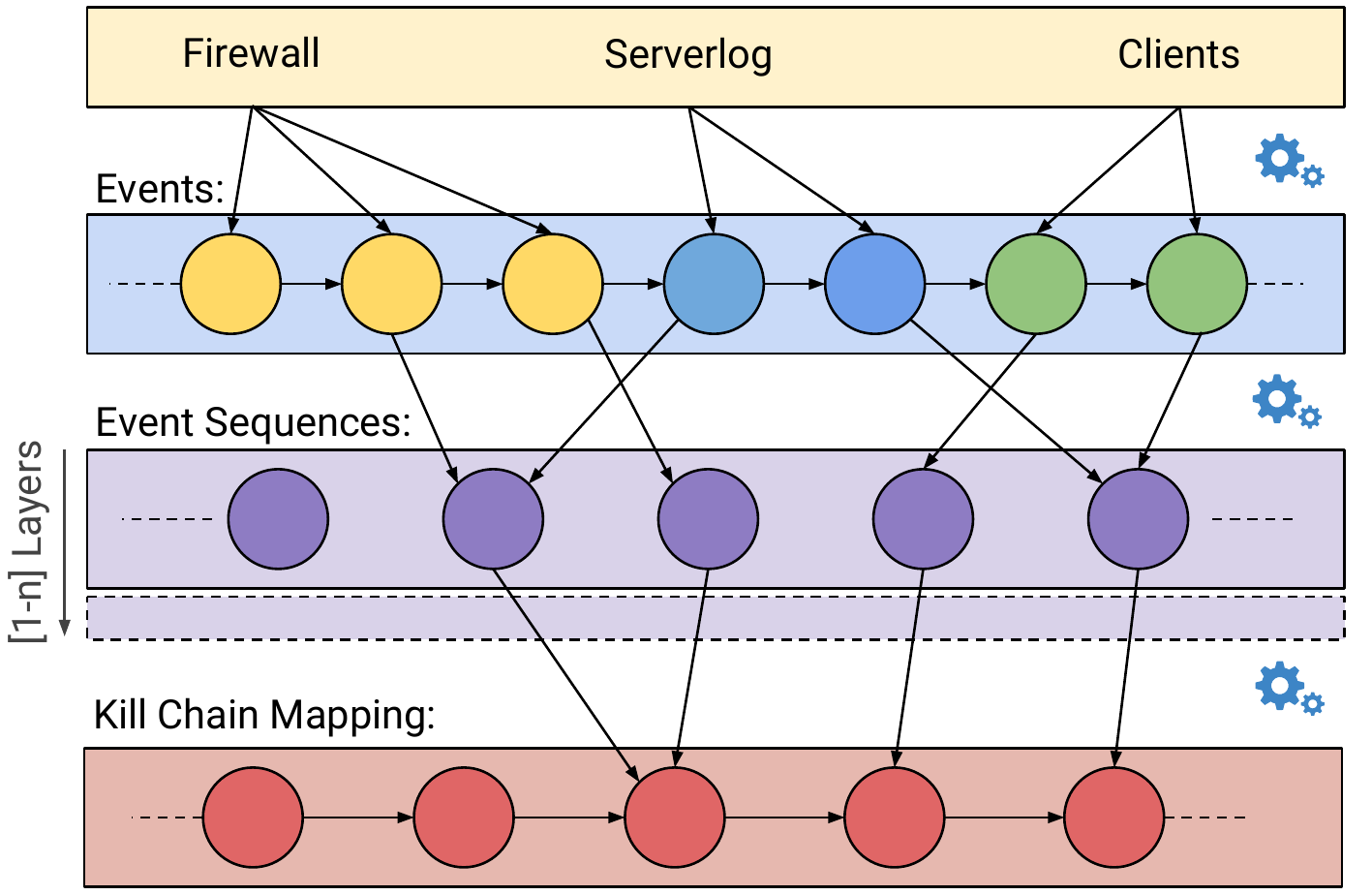}
\par
}
\caption{Detecting APT using Kill Chain Model in a Graph Database}
\end{figure}

To define different attacks or attack approaches, an adjusted kill chain is used.
In the kill chain process all steps of an adversary and its targets are described by a series of events.
The concept originates from US Department of Defense \cite{USDepartmentofDefense2007}, has been adjusted by Hutchins \cite{Hutchins2011} and refined by Uetz \cite{uezt2017} (see figure \ref{fig:Adjusted-Kill-Chain}).
Figure \ref{fig:Adjusted-Kill-Chain} shows the adjusted kill chain model\footnote{The kill chain elements ”1. Reconnaissance” and ”7. Prepare Exfiltration” are not displayed.} with deviating attack vectors\footnote{The green highlighted items are representing a specific APT and will be used later.}.
Every kill chain item represents a weak Indicator of Compromise (IoC), a signature which is used to detect a potentially compromised system \cite{fowler2016data}.
This makes it possible to identify different characteristics of APT attacks and react to different approaches like '2.1: Delivery with Spearphishing Email' and '2.2: Delivery with USB Flash Drive' (see fig. \ref{fig:Adjusted-Kill-Chain}).

\paragraph*{Graph-based Forensic Analysis}
The adjusted kill chain model has been implemented as directed graph.
Figure \ref{fig:Kill-Chain-Event-Mapping} shows the graph-database model where log events in chronological order representing the first layer (\emph{Events}, blue).
In most cases, an indicator is built of several log events.
Thus, a direct matching of log event to a kill chain element is not intended.
Between kill chain elements and log events, $n$ intermediate layers (\emph{event sequences}) are established (purple).
It is possible to create several layers of event sequences and also to use loops to illustrate the fact, that $n$ similar events can create a single event sequence.
If several corresponding event sequences match to the defined kill chain elements, an APT can be detected by analysing previous or following elements and their corresponding matches.
After a cyber attack is detected, it is possible to reconstruct the attack graph and find the security leak(s) and the corresponding victims.

\paragraph*{Anomaly Detection based on Machine Learning}
Besides graph analysis, statistical and machine learning methods are considered to identify attack vectors.
For this approach, in a first step windows log entries (e.g. login attempts or session times), firewalls and file audits are taken into account.
The SVM analysis is used to detect anomalies in the user behaviour evoked by APT processes on the users machine.
The collected data is processed by several SVMs and a profile based on the normal user behaviour is created.
The machine learning mechanism detects anomalies, based on the learned normal behaviour.

%
%
\section{Proof of Concept Case Study}
A specific APT is used to proof the detection concept:
\emph{Infecting a Windows system by executing malicious code located in an PDF file delivered by an infected spearphishing email, with the goal to exfiltrate all JPEG files.} 
For this purpose two sets of log data sources are used:
A security laboratory (SecLab) for students with about 13 workstations, a domain controller, firewall and proxy.
A framework to simulate corporate networks and cyber attacks with an arbitrary number of clients.

The logs of the above sources are aggregated with a state of the art commercial SIEM system\footnote{Prolog, NETZWERK Software GmbH}.
While following processing the log data gets pseudonymised, normalised and stored in a InfluxDB instance, a time series database and the main source for all analysis.
To be able to revoke the pseudonymisation, the framework uses a combination of a RSA cryptosystem and \cite{Shamir:1979:SS:359168.359176} secret sharing.

\paragraph*{Graph-based Forensic Analysis}
To be able to create log data of predefined attacks on a simulated company network and easily set the same conditions for repetitive runs of these attacks, the BREACH framework by Uetz et al. \cite{uezt2017} is used as one log source set.
The BREACH framework is being developed by FKIE\footnote{Fraunhofer-Institute for Communication, Information Processing and Ergonomics (FKIE), Bonn, Germany} in cooperation with OTH Regensburg.
The framework is capable to simulate various attack vectors and can also simulate normal user behaviour in a companies network fully controlled by the framework.
The simulator consists of a set of virtual machines simulating a medium sized enterprise network infrastructure with about 100 users.
All activities are logged and with the intrusion reconstruction every step of the attack can be identified.
Figure \ref{fig:Adjusted-Kill-Chain} shows the specific kill chain structure which is used in the case study (highlighted, green).
The detection of the example APT requires further analysis.

While the email is delivered to the client machine, the attacker starts a generic HTTP handler which runs in standby mode initially and waits for incoming connections.
The exploit \emph{Adobe Cooltype Sing} described in \cite{CVE-2010-2883} is used.
Through the handler further instructions can be delivered to the target.
In this case study the executed code searches for all JPEG image files on the local hard drive and send them to the attacker.

All log data will be stored in the InfluxDB.
While being processed with by graph database, new nodes are created for all involved machines and the algorithm matches the events to the affected sources.
This leads to a concatenated list of all events in a time period which are connected to the machines they are caused by.
The used algorithm for the graph analysis reveals if a kill chain has been fulfilled completely or just parts of it.
A complete kill chain indicates a successful breach by the APT.
Partly filled kill chains can be further investigated, if a similar attack vector is used, and are a first evidence for attacks.

\paragraph*{Anomaly Detection based on Machine Learning}
The anomaly detection with machine learning mechanisms is based on data of both data sources.
For the used SVMs and one-class SVMs both classifications, clean and malicious log data is needed.
While a normal SVM separates multiple predefined classes from each other, a one-class SVM is trained by the data of one class and separates it from the data  distinguishable from this class.
The feature vectors are based on file-audit, windows events (logon, logoff) and firewall logs.
For this malicious web traffic is generated with the BREACH framework coupled with manual malicious downloads in the SecLab.

%
%
\section{Evaluation}
Considering our case study the graph database has successfully created event sequences and linked the associated kill chain elements.
The shown attack involved the extraction of example confidential data, in this case JPEG files.
The amount of the extracted files is not known beforehand.
The algorithm managed to identify the attack correctly with a variable amount of confidential data.

\begin{figure}
\begin{center}
	\includegraphics[angle=90,width=0.85
	\textwidth]{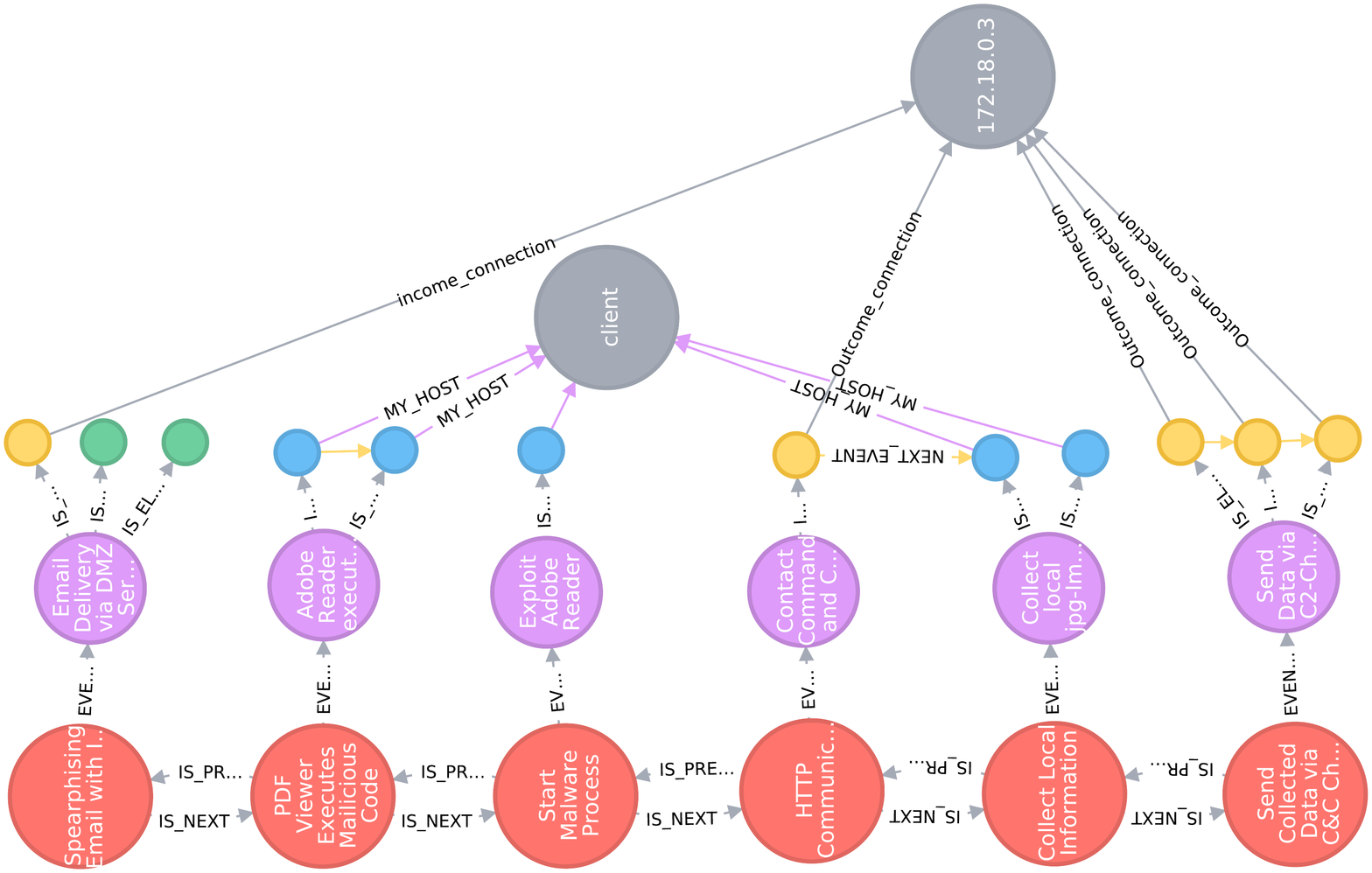}
	\captionof{figure}{Result of Graph Analysis for APT}
	\label{fig:graph-analysis-result}
\end{center}
\end{figure}

Figure \ref{fig:graph-analysis-result} shows the graphical result representation of the graph analysis in Neo4j\footnote{Figure \ref{fig:graph-analysis-result} only shows a simplified representation for demonstration propose.}.
The graph analysis successfully identifies the adversary (IP 172.18.0.3) of the simulation.
The example analysis is based on a simulated attack to one client with 1228 extracted log events (about 10 minutes of log data).
In a real world scenario much more log events even for a short period of time are created and need to be processed by the graph database.
With the second log source, the SecLab, it is possible to provide a realistic log data situation.

The research with the SVMs and one-class SVMs has shown a high accuracy from $95.33$\% to $98.67$\% when detecting anomalies in the log data.
It could be shown that in this case, the one-class SVMs had better results which can be traced back to a high amount of clean log data.
The data has shown, that there is a higher rate of detection in simulator data than in user created data which can be related to a more divergent behaviour of real world log data.
The SecLab data source is producing about 5 million events per week but simulations have shown that the SVM approach is also capable to analyse a much higher volume of log events.
To proof this, the data has been expanded with pseudo random data to proof the scalability.

%
%
\section{Conclusion and Future Work}
The purpose of this paper is to demonstrate that the use of a graph database is a practicable way for a forensic analysis of log data.
We have shown that the abstracted event sequence graph model for matching to a kill chain improves the automatic forensic analysis of log data.
To achieve this we have proposed an adjusted kill chain model as profound Indicator of Compromise.
To prove the theoretical concept, this framework was tested with an example attack model and successfully detects the breach.
We believe that this method is suitable for a variation of different APT attack vectors due to its flexibility.

The research with SVMs has already proven to be suitable to analyse massive log data and return a high detection rate of anomalies.
In the ongoing research we want to focus on  machine learning for automatic detection of attack patterns and the creation of new attack sequences.
For this approach we can use the different data sources for tagged attacks and clean log data.
In addition to that, we will implement a fraud detection with fraud rings.
This approach will help to find connections between affected systems and to identify potential additional breach victims.
One major challenge to overcome is the high variety of log data from different sources.
To achieve a fast and reliable framework we want to normalise and enrich the log data with additional information.

\printbibliography if you use biblatex/Biber
\end{document}